\documentclass[prd,,aps,nofootinbib,amsmath,amssymb]{revtex4}
\usepackage{amssymb,amsmath,epsfig,bm}
\usepackage[usenames,dvipsnames]{color}
\usepackage[bookmarks]{hyperref}

\newtheorem{lemma}{Lemma}
\newtheorem{proposition}{Proposition}
\newtheorem{theorem}{Theorem}

\newcommand{\proof}{\noindent {\bf Proof. }}

\newcommand{\qed}{\hfill $\fbox{\hspace{0.3mm}}$ \vspace{.3cm}} 

\newcommand{\Real}{\mathbb{R}}

\begin{document}

\title{Cauchy horizon stability in a collapsing spherical dust cloud I: geometric optic approximation and spherically symmetric test fields}
\author{N\'estor Ortiz and Olivier Sarbach}
\affiliation{Instituto de F\'isica y Matem\'aticas, Universidad Michoacana de San Nicol\'as de Hidalgo, Edifico C-3, Ciudad Universitaria, 58040, Morelia, Michoac\'an, M\'exico.}
\email{nestor@ifm.umich.mx and sarbach@ifm.umich.mx}

\begin{abstract}
A spherical dust cloud which is initially at rest and which has a monotonously decaying density profile collapses and forms a shell-focussing singularity. Provided the density profile is not too flat, meaning that its second radial derivative is negative at the center, this singularity is visible to local, and sometimes even to global observers. According to the strong cosmic censorship conjecture, such naked singularities should be unstable under generic, nonspherical perturbations of the initial data or when more realistic matter models are considered. 

In an attempt to gain some understanding about this stability issue, in this work we initiate the analysis of a simpler but related problem. We discuss the stability of test fields propagating in the vicinity of the Cauchy horizon associated to the naked central singularity. We first study the high-frequency limit and show that the fields undergo a blueshift as they approach the Cauchy horizon. However, in contrast to what occurs at inner horizons of black holes, we show that the blueshift is uniformly bounded along incoming and outgoing null rays. Motivated by this boundedness result, we take a step beyond the geometric optic approximation and consider the Cauchy evolution of spherically symmetric test scalar fields. We prove that under reasonable conditions on the initial data a suitable rescaled field can be continuously extended to the Cauchy horizon. In particular, this result implies that the physical field is everywhere finite on the Cauchy horizon away from the central singularity.
\end{abstract}

\maketitle

\section{Introduction}
\label{Sec:Intro}

The simplest model describing the gravitational collapse of a massive body in general relativity is the one of Tolman-Bondi (TB), which describes the total collapse of a spherical dust cloud (see, for instance, Ref.~\cite{MTW-Book}). Although the underlying assumptions of spherical symmetry and zero pressure are hardly satisfied in a realistic collapse, the advantage of this model relies in the fact that it can be described by closed-form expressions for the spacetime metric and energy density when co-moving, synchronous coordinates are employed. This facilitates the analysis of the physical properties of the model, which has already played an important role in the understanding of black hole formation in the seminal work by Oppenheimer and Snyder~\cite{jOhS39}. Under rather reasonable assumptions on the initial data, namely the initial density and velocity profiles, one can show that a shell-focusing singularity forms, where the density and hence also the curvature blow up. Despite the fact that the metric is known in explicit form, the analysis of the causal structure in the vicinity of the shell-focusing singularity is a nontrivial task since the understanding of the behaviour of the light rays (even those with zero angular momentum) requires the solution of singular, nonlinear ordinary equations. The first systematic study of the radial light rays has been undertaken by Christodoulou~\cite{dC84}. Interestingly, his work (and also earlier numerical work by Eardley and Smarr~\cite{dElS79}) revealed a picture that is rather different than the one obtained from the simple Oppenheimer-Snyder scenario of homogeneous density. Indeed, the analysis in~\cite{dC84} showed that for regular, time-symmetric generic initial data there exist infinitely many radial light rays emanating from the central singularity, implying that the later is visible to local observers. Furthermore, it was shown in~\cite{dC84} that depending on the initial density profile, some of these light rays arrive at the surface of the cloud before the horizon forms, so that they extend all the way to null infinity. In this case the shell-focusing singularity is ``globally naked'' in the sense that it is visible to observers that are located arbitrarily far from the dust cloud. For generalizations of these results to time-asymmetric initial data, see Refs.~\cite{rN86,iDpJ92,Joshi-Book}. For a self-contained exposition of these results we refer to our recent work in~\cite{nOoS11}, where we also provide a sufficient condition on the initial data for the occurrence of a globally naked singularity and present an algorithm for numerically generating conformal diagrams.

In this article we initiate a detailed analysis of test fields propagating on the fixed (but dynamical) spacetime geometry given by the TB collapse model. We are particularly interested in the behaviour of such test fields in a vicinity of the central naked singularity and the Cauchy horizon, corresponding to the first light ray emanating from the central singularity. Assuming regular initial data for the test field on a Cauchy surface, a relevant question is whether or not the field or its energy density grow arbitrarily large when they approach the Cauchy horizon. Such a divergent behaviour would suggest that the Cauchy horizon is unstable when the self-gravity of the field is taken into account. Indeed, such an instability was found at the inner horizon of a Reissner-Nordstr\"om black hole, where early numerical work~\cite{mSrP73} evolving test fields indicated an instability of the Cauchy (inner) horizon, a fact that was later confirmed by self-consistent calculations~\cite{ePwI90,mD05} resulting in the famous mass inflation scenario.

There has been some previous work regarding the stability of the Cauchy horizon associated to naked singularities in the TB dust collapse model. In the geometric optic approximation of test fields, Christodoulou~\cite{dC84} showed that, for time-symmetric initial data, a radial light ray emitted from the center of the cloud -even close to the singularity- does not undergo an infinite redshift. Based on a similar approximation, Waugh and Lake~\cite{bWkL89} studied the stability of the Cauchy horizon for the particular case of self-similar collapse and did not find evidence for an instability. Duffy and Nolan took a step beyond the geometric limit approximation and studied the stability of the Cauchy horizon in the case of self-similar, marginally bound collapse under odd-~\cite{eDbN11b} and even-parity~\cite{eDbN11} linear metric perturbations. In the odd-parity sector, they found that the perturbations remain finite at the Cauchy horizon. In contrast, in the even-parity sector, they showed that the perturbations diverge on the Cauchy horizon. Other progress concerning the marginally bound case include the work by Iguchi, Harada and Nakao~\cite{hItHkN98} who analyzed numerically the stability of the Cauchy horizon under odd-parity linear gravitational perturbations. When the matter perturbations vanish, they show that the metric perturbations do not diverge at the Cauchy horizon. On the other hand, they also found that the metric perturbations diverge when the matter perturbations are not zero~\cite{hItHkN99}. Nevertheless, this divergence occurs only at the central singularity and does not propagate along the Cauchy horizon. The same authors extended their numerical studies to even-parity perturbations and concluded that they diverge near the Cauchy horizon while the energy flux keeps bounded~\cite{hItHkN00}, suggesting that the singularity is not a strong source of gravitational waves.

The main purpose of our work is to derive rigorous results in order to obtain a deeper insight into the physical effects that occur near the naked central singularity and the Cauchy horizon. For this reason, we analyze the generic TB collapse, without restricting ourselves to particular cases as marginally boundedness or self-similarity. The remaining of this article is organized as follows. In the next section we review the basic properties of the TB collapse model, state our assumptions on the initial data, recall important results on the propagation of radial light rays emanating from the central singularity and state some preliminary results that are relevant for the analysis that follows. Next, in section~\ref{Sec:BlueShift} we compute the gravitational redshift factor for a light ray that is emitted and received by two observers which are co-moving with the dust shells. We first show that for observers which are sufficiently close to the central singularity, this factor is always negative, implying a gravitational \emph{blueshift}, even along outgoing null rays. As we show, the fact that there is a blueshift and not a redshift is an effect which is due to the collapse of the cloud, implying that outgoing photons move towards a region of \emph{stronger} gravity. As the path of the light ray moves closer and closer to the central singularity, the blueshift becomes more pronounced, as expected. However, we show in section~\ref{Sec:BlueShift} that the total blueshift along such a light ray is uniformly bounded. Motivated by these considerations, in section~\ref{Sec:SphSym} we analyze the dynamics of a spherically symmetric scalar field $\Phi$ on the TB background. The propagation of such a field is described by an effective, two-dimensional wave equation with potential $V$ for the rescaled field $\psi = r\Phi$, where $r$ denotes the areal radius. We prove that under a suitable integrability condition on the effective potential $V$, the rescaled field $\psi$ can be continuously extended to the Cauchy horizon. Then, we verify that this condition on $V$ is satisfied and conclude that the physical field $\Phi$ must be bounded by a constant divided by $r$ in the vicinity of the Cauchy horizon. This implies, in particular, that $\Phi$ is finite everywhere on the Cauchy horizon away from the central singularity. Therefore, if $\Phi$ should diverge at the central singularity, this divergence cannot propagate along the Cauchy horizon. A discussion of our results and an outlook are presented in section~\ref{Sec:Conclusions}. More general results concerning the propagation of linear and nonlinear test fields in the vicinity of the Cauchy horizon of a TB dust collapse spacetime will be presented elsewhere.

\section{Tolman-Bondi collapse: review, preliminaries and notation}
\label{Sec:Model}

In this section we first review the TB collapse model, which describes the complete gravitational collapse of a spherical dust cloud. Then, we summarize our assumptions on the initial data. These assumptions are taken from our earlier work~\cite{nOoS11} and essentially impose physically ``reasonable'' conditions on the initial density and velocity profiles. Furthermore, our conditions guarantee that no singularities occur at a finite radius, therefore avoiding shell-crossing singularities. After stating our assumptions we review the results in~\cite{nOoS11} regarding the propagation of radial light rays which are relevant for this work. Finally, we introduce some notation and derive some preliminary results used later in this article.

The spherically symmetric solutions of the field equations describing a self-gravitating dust configuration can be explicitly parametrized in terms of co-moving, synchronous coordinates $(\tau,R,\vartheta,\varphi)$. Here, $R=const.$ describes the world surfaces of the collapsing dust shells, where the label $R$ is chosen such that it coincides with the shells' areal radius at initial time $\tau=0$. $\tau$ is the proper time measured by a radial observer moving along a collapsing dust shell, and $(\vartheta,\varphi)$ are standard polar coordinates on the invariant two-spheres. The metric is determined by the function $r(\tau,R)$ which describes the areal radius at the event $(\tau,R,\vartheta,\varphi)$. Therefore, for fixed $R$, the function $\tau\mapsto r(\tau,R)$ describes the evolution of the dust shell labeled by $R$, and according to the definition of $R$, $r(0,R) = R$. As a consequence of Einstein's field equations (see, for example,~Ref.~\cite{MTW-Book}), this function satisfies the following one-dimensional mechanical system,
\begin{equation}
\frac{1}{2} \dot{r}(\tau,R)^2 + V(r(\tau,R), R) = E(R),\qquad
V(r,R) := -\frac{m(R)}{r},
\label{Eq:1DMechanical}
\end{equation}
where the dot denotes partial derivative with respect to $\tau$, and where $m(R)$ denotes the Misner-Sharp mass function~\cite{cMdS64}.\footnote{$m(R)$ is also the Hawking mass~\cite{sH68} associated to the invariant two-spheres.} The mass and energy profiles $m(R)$ and $E(R)$ are determined by the initial density and velocity profiles $(\rho_0(R),v_0(R))$ according to
\begin{displaymath}
m(R) = 4\pi G\int\limits_0^R \rho_0(\bar{R})\bar{R}^2 d\bar{R},\qquad
E(R) = \frac{1}{2} v_0(R)^2 - \frac{m(R)}{R},
\end{displaymath}
with Newton's constant $G$. Once the function $r(\tau,R)$ is known, the spacetime metric ${\bf g}$, the four-velocity ${\bf u}$ and the energy density $\rho$ are obtained by means of the following explicit formulae:
\begin{eqnarray}
{\bf g} &=& -d\tau^2 + \frac{dR^2}{\gamma(\tau,R)^2}
 + r(\tau,R)^2(d\vartheta^2 + \sin^2\vartheta\, d\varphi^2),\qquad
\gamma(\tau,R) := \frac{\sqrt{1 + 2E(R)}}{r'(\tau,R)},
\label{Eq:MetricSol}\\
{\bf u} &=& \frac{\partial}{\partial\tau}\; , \qquad
\rho(\tau,R) = \rho_0(R)\left( \frac{R}{r(\tau,R)} \right)^2\frac{1}{r'(\tau,R)},
\label{Eq:FluidSol}
\end{eqnarray}
where the prime denotes partial differentiation with respect to $R$. It is convenient to introduce the two functions
\begin{displaymath}
c(R) := \frac{2m(R)}{R^3},\qquad
q(R) := \sqrt{E(R)/V(R,R)} = \sqrt{1 - \frac{R v_0(R)^2}{2m(R)}},
\end{displaymath}
describing (up to a numerical factor) the mean density within the dust shell $R$ and the square root of the ratio between the total and initial potential energy. In terms of these quantites, the assumptions in~\cite{nOoS11} can be summarized as follows:
\begin{enumerate}
\item[(i)] $\rho_0$ and $v_0$ have even and odd $C^\infty$-extensions, respectively, on the real axis (regular, smooth initial data),
\item[(ii)] $\rho_0(R) > 0$ for all $0\leq R < R_1$ and $\rho_0(R) = 0$ for $R\geq R_1$ (finite, positive density cloud)
\item[(iii)] $c'(R)\leq 0$ for all $R > 0$ (monotonically decreasing mean density),
\item[(iv)] $2m(R)/R < 1$ for all $R > 0$ (absence of trapped surfaces on the initial slice),
\item[(v)] $v_0(R)/R < 0$ for all $R\geq 0$ (collapsing cloud),
\item[(vi)] $(v_0(R)/R)^2 < 2m(R)/R^3$  for all $R\geq 0$ (bounded collapse),
\item[(vii)] $q'(R)\geq 0$ for all $R > 0$ (exclusion of shell-crossing singularities),
\item[(viii)] For all $R\geq 0$, we have $q'(R)/R > 0$ whenever $c'(R)/R = 0$ (non-degeneracy condition).
\end{enumerate}
Notice that condition (i) ensures that the functions $c$ and $q$ have even $C^\infty$-extensions on the real axis, and that condition (vi) implies that $q(R) > 0$ for all $R\geq 0$. Condition (viii) implies the existence of a null portion of the singularity which is visible at least to local observers. Explicit four-parameter families of initial data $(\rho_0(R),v_0(R)$ satisfying all of these conditions except (i) have been constructed in Ref.~\cite{nOoS11} (see Eq. (28)) and in Ref.~\cite{nO12} (see Eq.~(4)). In fact, these families also satisfy condition (i) except at the surface of the cloud $R = R_1$, where $\rho$ and $v_0$ are only continuous. The lack of smoothness at the surface of the cloud does not affect our results below, which mostly refer to regions close to the central singularity.

Under the conditions (i)--(viii) the solution of the mechanical system~(\ref{Eq:1DMechanical}) is given by the explicit formula
\begin{equation}
r(\tau,R) 
 = \frac{R}{q(R)^2} \left[f^{-1}\left( f(q(R)) + \sqrt{c(R)}q(R)^3\tau \right) \right]^2,
\label{Eq:Sol}
\end{equation}
where $f$ is the strictly decreasing function
\begin{equation}
f: [0,1] \to [0,\pi/2],\quad x\mapsto x\sqrt{1 - x^2} + \arccos(x),
\label{Eq:fDef}
\end{equation}
which is $C^\infty$-differentiable on the interval $[0,1)$ and whose first derivative is  $f'(x) = -2x^2/\sqrt{1 - x^2}$, $0\leq x < 1$. The function $r(\tau,R)$ is well-defined for all $R\geq 0$ and $0\leq\tau < \tau_s(R)$, where the boundary $\tau = \tau_s(R)$ parametrizes the location of the shell-focusing singularity, for which $r/R$ vanishes. From Eq.~(\ref{Eq:Sol}) one obtains
\begin{equation}
\tau_s(R) = \frac{\frac{\pi}{2} - f(q(R))}{\sqrt{c(R)}q(R)^3},\qquad R\geq 0.
\label{Eq:taus}
\end{equation}
Since the energy density $\rho$ diverges, Einstein's field equations imply that the Ricci scalar diverges at the shell-focusing singularity. Therefore, the boundary points $\tau = \tau_s(R)$ represent a curvature singularity. The tidal forces are much stronger near such points than in the case of shell-crossing singularities~\cite{pSaL99}. Outside the cloud, $R > R_1$, where $\rho_0=0$, the spacetime is isometric to a subset of the Schwarzschild-Kruskal manifold according to Birkhoff's theorem, see for example Ref.~\cite{Straumann-Book}.

For the following, it is convenient to use new local coordinates $(y,R)$ instead of $(\tau,R)$, with $y$ defined by $y:=\sqrt{r(\tau,R)/R}$. As shown in \cite{nOoS11} this facilitates the analysis of the light rays in several ways. First, in these coordinates the spacetime region inside the collapsing dust cloud is the rectangular region $(y,R)\in (0,1)\times (0,R_1)$, with the initial surface and the singularity corresponding to the lines $y=1$ and $y=0$, respectively. Second, using $y$ instead of $\tau$, allows one to get around computing the inverse of the function $f$ defined in Eq.~(\ref{Eq:fDef}). The equation for the outgoing radial null geodesics in these coordinates is 
\begin{equation}
\frac{dy}{dR} = \frac{1}{2}\sqrt{1-q(R)^2y^2}\left[ \frac{R\Lambda(y,R)}{y^2}
\left( 1 - \frac{R Q(R)}{ y}\sqrt{1 - q(R)^2y^2} \right) - Q(R) \right],
\label{Eq:dy/dR}
\end{equation}
where the functions $\Lambda: [0,1)\times [0,R_1] \to \Real$, $Q: [0,R_1]\to \Real$, and $g,h: (0,1)\times [0,1)\to\Real$ are defined as
\begin{eqnarray*}
\Lambda(y,R) &:=& 2\frac{q'(R)}{Rq(R)} h(q(R),y) 
  - \frac{c'(R)}{2Rc(R)} g(q(R),y),\\
Q(R) &:=& \sqrt{\frac{c(R)}{1 - R^2 q(R)^2 c(R)}},\\
g(q,y) &:=& \frac{ f(qy) - f(q) }{q^3},\\
h(q,y) &:=& \frac{1}{\sqrt{1-q^2}} - \frac{y^3}{\sqrt{1- q^2y^2}} - \frac{3}{2}g(q,y).
\end{eqnarray*}
It follows from Lemma 1 of Ref.~\cite{nOoS11} that these functions are strictly positive and $C^\infty$-differentiable on their domain. Before we proceed, let us mention two limiting cases which can be included in our analysis below with the following adjustments:
\begin{enumerate}
\item $q=1$ (time-symmetric initial data)\\
This case corresponds to zero initial velocity, and here $g(q,y) = f(y)$, while the function $\Lambda$ is
\begin{displaymath}
\Lambda(y,R) = -\frac{c'(R)}{2Rc(R)} f(y),
\end{displaymath}
while the function $h$ is void.
\item $q=0$ (marginally bound case)\\
This corresponds to the case where each shell has zero energy, $E(R)=0$. Here, the functions $g(q,y)$ and $\Lambda(y,R)$ have the following expressions:
\begin{displaymath}
g(q,y) = \frac{2}{3}(1-y^3),\qquad
\Lambda(y,R) = -\frac{c'(R)}{3R c(R)} (1-y^3).
\end{displaymath}
\end{enumerate}

\begin{figure}[h!]
\begin{center}
\includegraphics[width=9.4cm]{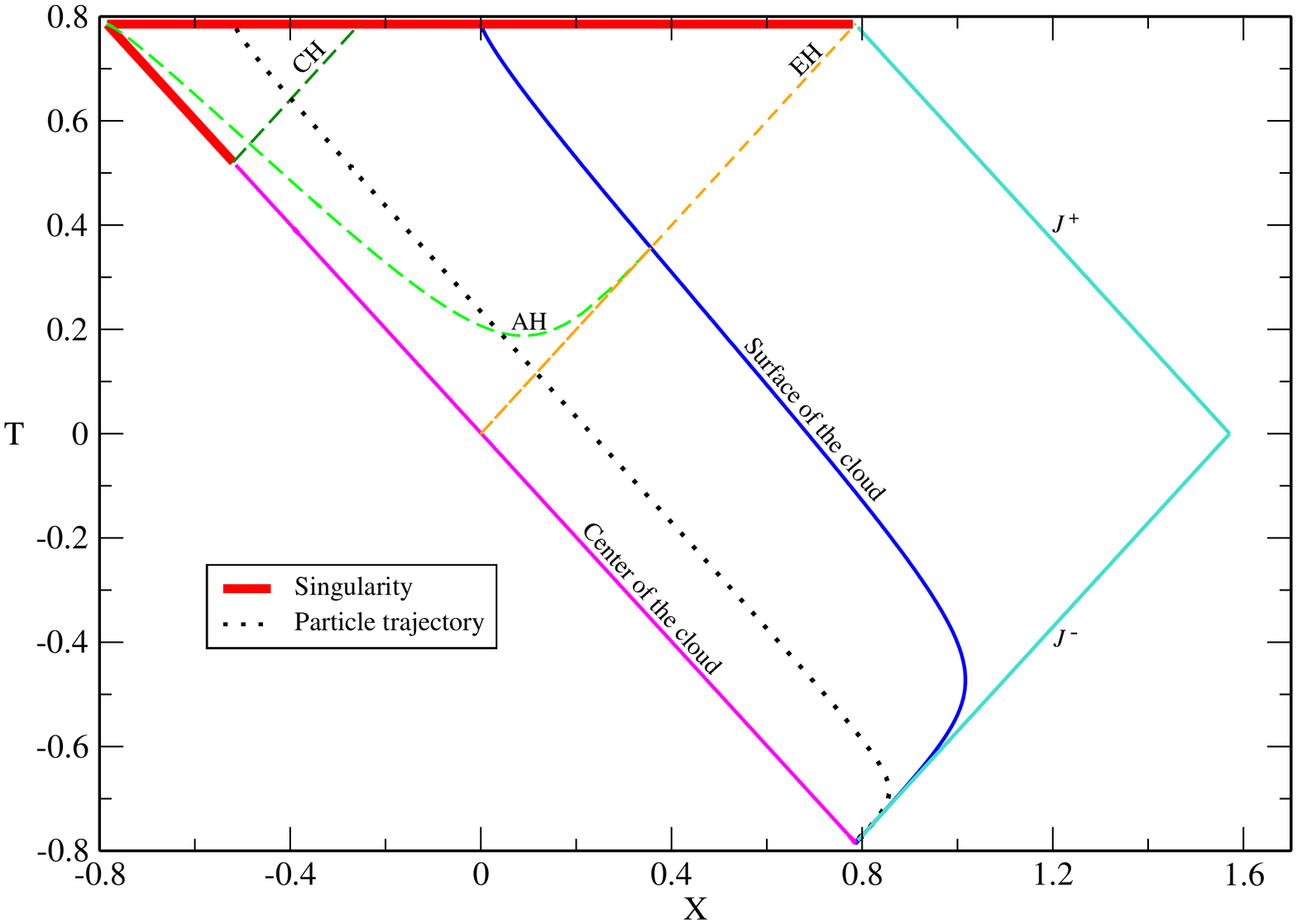}
\hspace{-1.2cm}
\includegraphics[width=9.4cm]{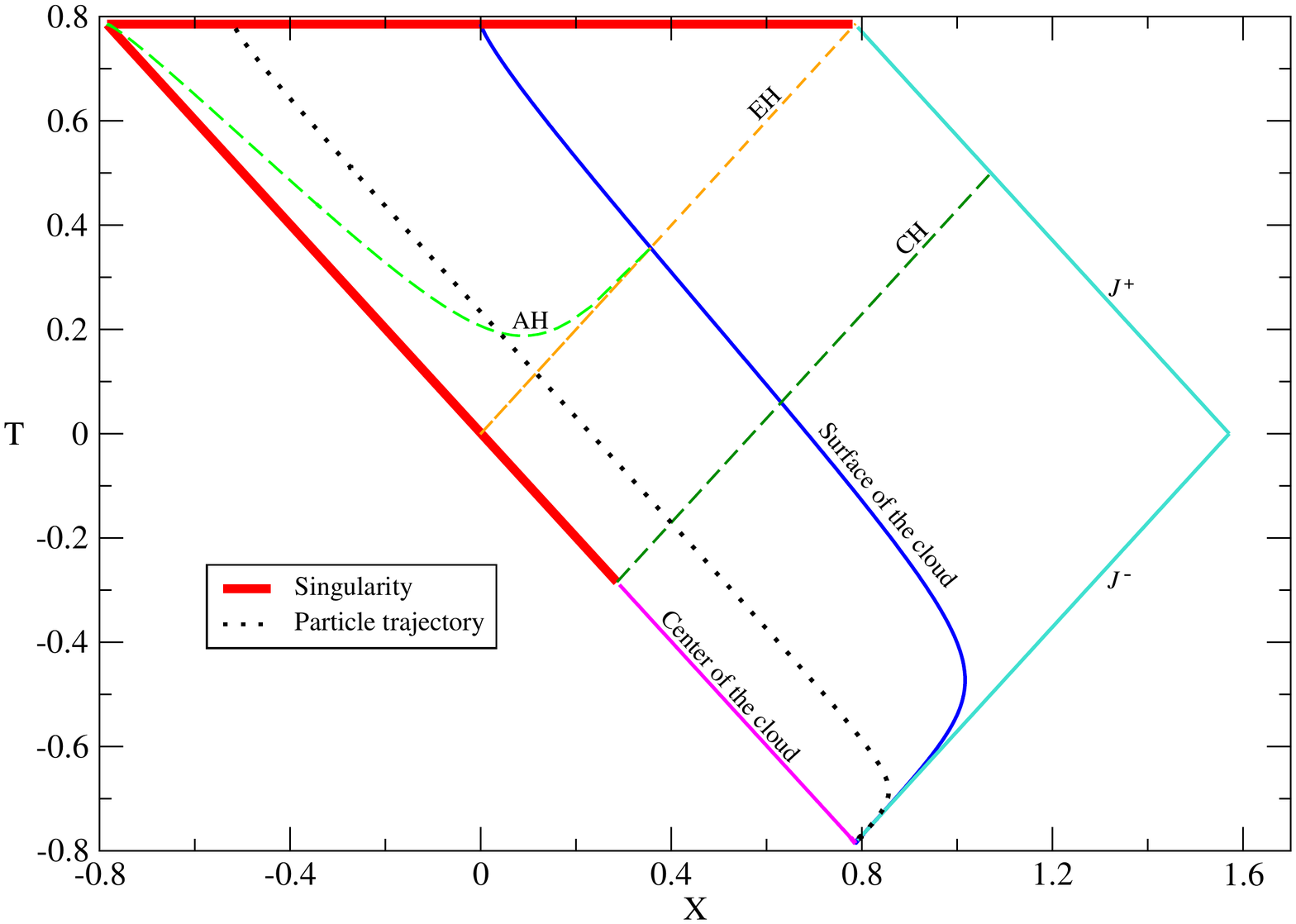}
\end{center}
\caption{\label{Fig:Hidden-Naked} Spacetime diagrams in conformal coordinates $(T,X)$ describing a TB dust collapse with the initial data given in Eq.~(28) of Ref.~\cite{nOoS11}. Left panel: The singularity is hidden inside the black hole region. Right panel: A portion of the null singularity is globally naked. The lines denoted by ``AH'', ``EH'' and ``CH'' refer to the apparent, event and Cauchy horizons, respectively. Details on the numerical construction of these diagrams are given in Ref.~\cite{nOoS11}. The parameters used correspond to the ones described in figures~3~and~4 of that reference.}
\end{figure}

In terms of the new coordinates $(y,R)$ the metric coefficient $\gamma$ in Eq.~(\ref{Eq:MetricSol}) and the energy density $\rho$ given in Eq.~(\ref{Eq:FluidSol}) can be computed according to the following formulae:
\begin{eqnarray}
\gamma(y,R) &=& \frac{\sqrt{1 - R^2 q(R)^2 c(R)}}{r'(y,R)},
\label{Eq:gamma}\\
\rho(y,R) &=& \frac{\rho_0(R)}{y^4 r'(y,R)},
\label{Eq:rho}
\end{eqnarray}
where $\rho_0(R) = 2m'(R)/(8\pi G R^2) = [R^3 c(R)]'/(8\pi G R^2)$ and the partial derivative of $r(\tau,R)$ with respect to $R$ is
\begin{equation}
r'(y,R) = y^2\left( 1 + \frac{R^2}{y^3}\sqrt{1 - q(R)^2y^2} \Lambda(y,R) \right).
\label{Eq:rprime}
\end{equation}
In addition, if follows from Eq.~(\ref{Eq:Sol}) that the proper time at $(y,R)$ is
\begin{equation}
\tau(y,R) = \frac{g(q(R),y)}{\sqrt{c(R)}}. 
\label{Eq:tau}
\end{equation}
The causal structure of the spacetime described by Eqs.~(\ref{Eq:MetricSol},\ref{Eq:FluidSol}) has been analyzed in Refs.~\cite{dC84,rN86}, and more recently in Ref.~\cite{nOoS11}, where conformal diagrams are constructed based on local analysis and numerical integration of Eq.~(\ref{Eq:dy/dR}). The common feature is the presence of a locally naked central singularity and an associated Cauchy horizon that emanates from it. Depending on whether or not the Cauchy horizon intersects the surface of the cloud before or after the apparent horizon, the central singularity is globally naked or hidden inside a black hole, see figure~\ref{Fig:Hidden-Naked} for two examples taken from~\cite{nOoS11}.

\begin{figure}[h!]
\begin{center}
\includegraphics[width=9cm]{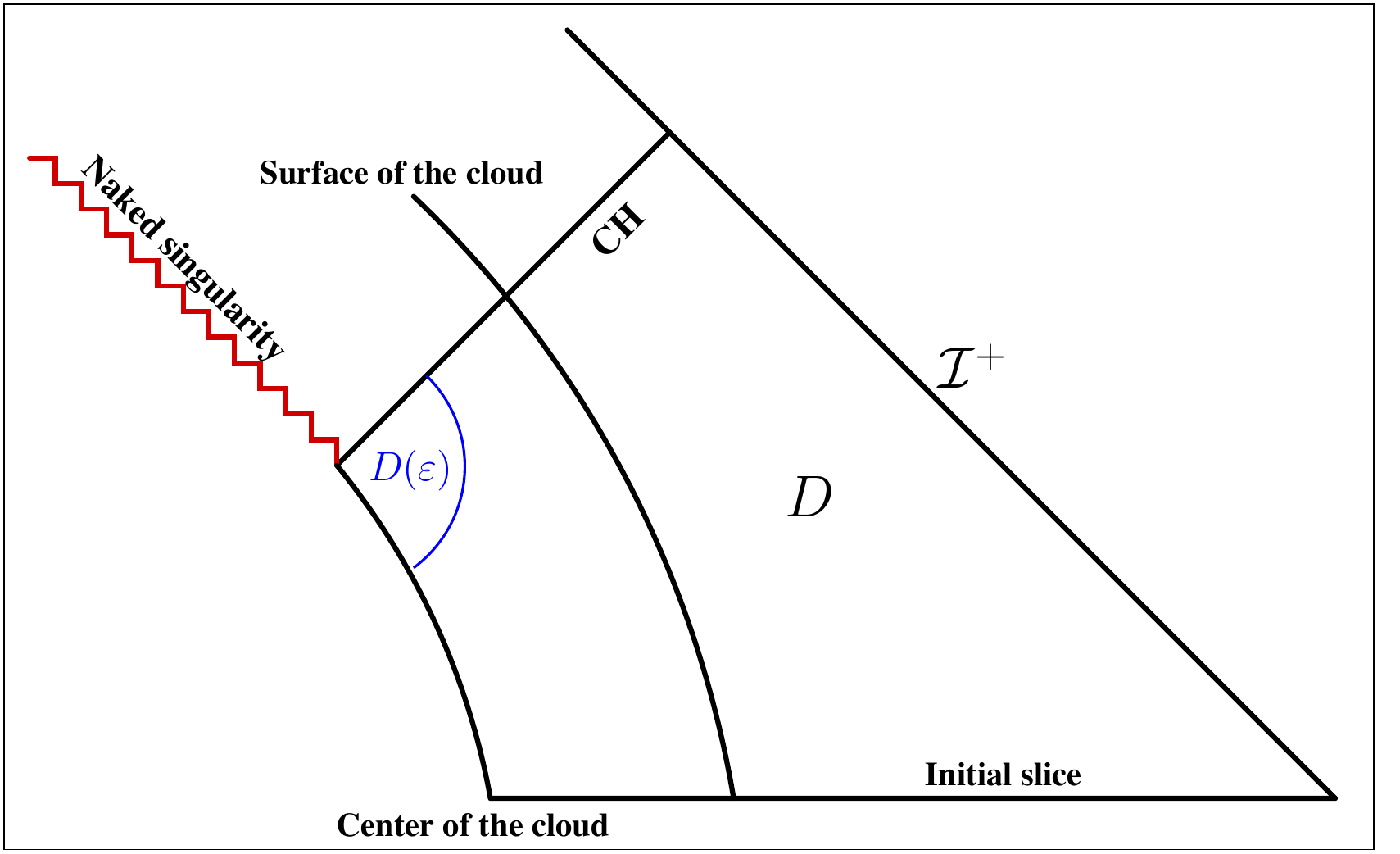}
\end{center}
\caption{\label{Fig:D_epsilon} Conformal diagram illustrating the maximal development $D$ of the initial slice and the small subsets $D(\varepsilon)$ close to the central singularity. Here ``CH'' denotes the Cauchy horizon.}
\end{figure}

In the following, for definiteness we will only consider the case of a globally naked singularity, although most of our results also hold in the black hole case. We will focus our attention to the region $D$ of spacetime describing the maximal development of the initial surface $\tau=0$, see figure~\ref{Fig:D_epsilon}. In terms of the coordinates $(y,R)$ this region is described by
\begin{displaymath}
D := \{ (y,R) : R\geq 0, y_{CH}(R) < y \leq 1 \},
\end{displaymath}
where the curve $y_{CH}(R)$ describes the Cauchy horizon, that is, the first light ray emanating from the singularity. For $0 < R < \delta$ small enough it was shown in Proposition 2 of Ref.~\cite{nOoS11} that $y_{CH}(R)$ has the following form:
\begin{displaymath}
y_{CH}(R)^3 = \frac{3\Lambda_0}{4} R^2\left[ 1 + z(R^{1/3}) \right],\qquad
R\in [0,\delta),
\end{displaymath}
where $\Lambda_0 := \Lambda(0,0) > 0$ and $z: [0,\delta)\to\Real$ is a $C^\infty$-function satisfying $z(0)=0$. Besides the spacetime region $D$, we will also consider for each $\varepsilon \in (0,1)$ the small spacetime regions $D(\varepsilon)\subset D$ near the central singularity $(0,0)$ defined as
\begin{equation}
D(\varepsilon) := \{ (y,R) : 0 \leq R \leq R(\varepsilon), y_{CH}(R) < y \leq y(\varepsilon) \},
\label{Eq:DepsDef}
\end{equation}
where the functions $R(\varepsilon) \in (0,\delta)$ and $y(\varepsilon) := y_{CH}(R(\varepsilon))$ are chosen such that they converge to zero when $\varepsilon\to 0$ and such that
\begin{enumerate}
\item[(i)] $z(R^{1/3}) \geq -\varepsilon$ for all $0\leq R \leq R(\varepsilon)$
\item[(ii)] $\Lambda(y,R)\leq (1 + \varepsilon)\Lambda_0$ for all $0\leq R\leq R(\varepsilon)$ and $0\leq y\leq y(\varepsilon)$,
\end{enumerate}
see figure~\ref{Fig:D_epsilon}.

It follows for each $(y,R)\in D(\varepsilon)$ that
\begin{equation}
0\leq \frac{R^2}{y^3}\sqrt{1 - q(R)^2y^2} \Lambda(y,R)
\leq \frac{R^2}{y_{CH}(R)^3}\Lambda(y,R)
= \frac{4}{3}\frac{1}{1 + z(R^{1/3})}\frac{\Lambda(y,R)}{\Lambda_0} 
\leq \frac{4}{3}\frac{1 + \varepsilon}{1 - \varepsilon}.
\label{Eq:R2y3Estimate}
\end{equation}
This and Eq.~(\ref{Eq:rprime}) imply that inside the region  $D(\varepsilon)$,
\begin{equation}
y^2\leq r'(y,R) \leq \left( 1 + \frac{4}{3}\frac{1 + \varepsilon}{1 - \varepsilon} \right)y^2.
\label{Eq:y2Estimat}
\end{equation}
In view of Eqs.~(\ref{Eq:gamma},\ref{Eq:rho}) this yields the following behaviour for the metric coefficient $\gamma$ and the energy density $\rho$ in the vicinity of the central singularity:

\begin{lemma}
Let $\varepsilon\in (0,1)$. Then there are constants $m < M$ such that for all $(y,R)\in D(\varepsilon)$,
\begin{equation}
\frac{m}{y^2} \leq \gamma(y,R) \leq \frac{M}{y^2},\qquad
\frac{m}{y^6} \leq \rho(y,R) \leq \frac{M}{y^6}.
\end{equation}
\end{lemma}
Therefore, $\gamma$ diverges as $1/y^2$ and $\rho$ as $1/y^6$ as the singularity is approached from within the maximal development $D$ of the initial data surface. In the next two sections we analyze the propagation of test fields on the background spacetime $(D,{\bf g})$ and show that despite these divergences, the fields do not behave ``too badly'' as the central singularity $(0,0)$ and the Cauchy horizon are approached.

In the following, the radial vector fields ${\bf u}$, ${\bf w}$, ${\bf k}$ and ${\bf l}$ play an important role. ${\bf u} = \partial_\tau$ is the four-velocity of the radial observers co-moving with the collapsing dust shells, see Eq.~(\ref{Eq:FluidSol}), ${\bf w} := \gamma(\tau,R)\partial_R$ is the unit outward radial vector orthogonal to ${\bf u}$, and ${\bf k} := {\bf u} + {\bf w}$, ${\bf l} := {\bf u} - {\bf w}$ are radial out- and ingoing null vector fields.

\section{Boundedness of the blueshift of light rays and null dust}
\label{Sec:BlueShift}

\begin{figure}[h!]
\begin{center}
\includegraphics[height=4.6cm]{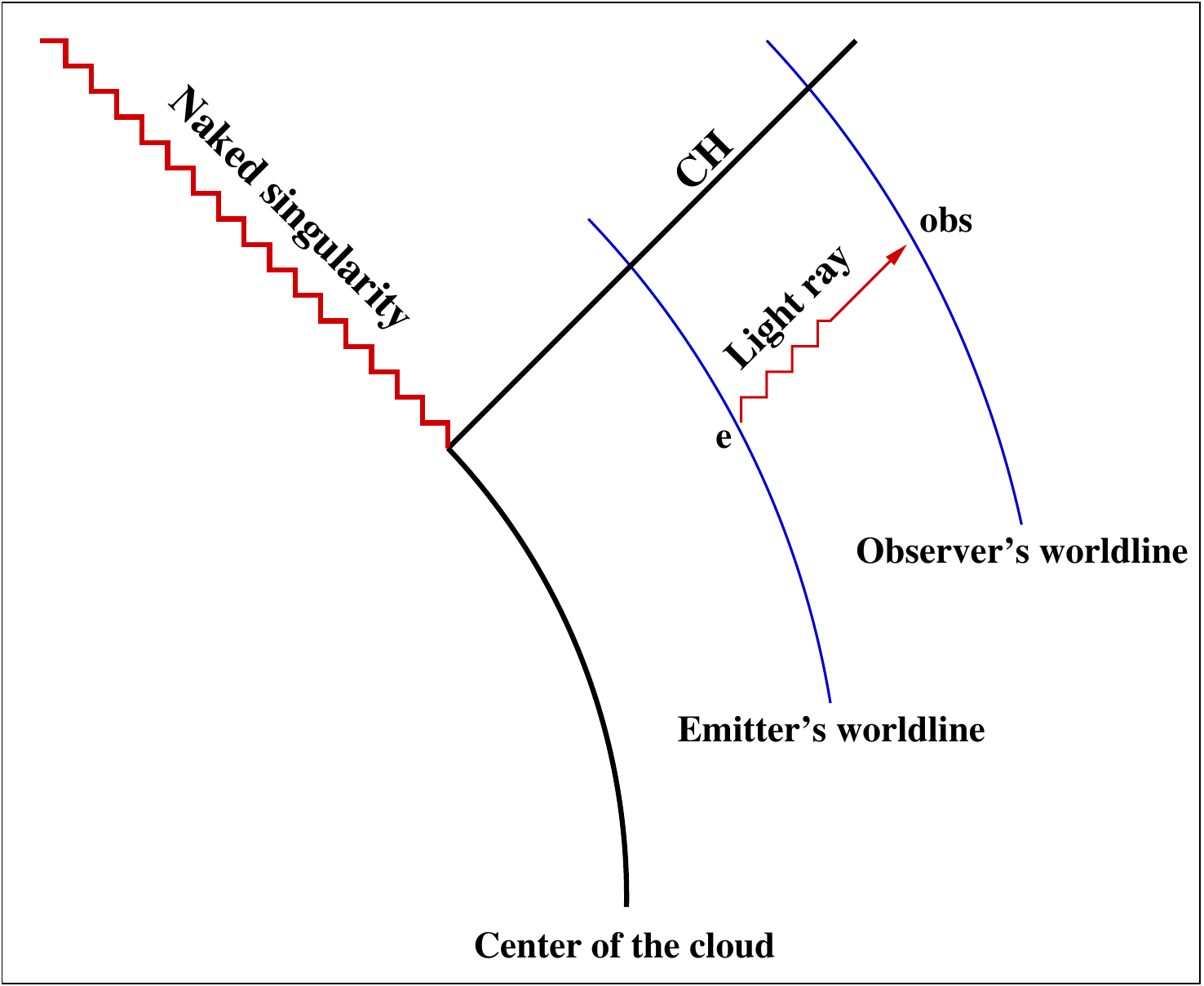}
\hspace{0.3cm}
\includegraphics[height=4.6cm]{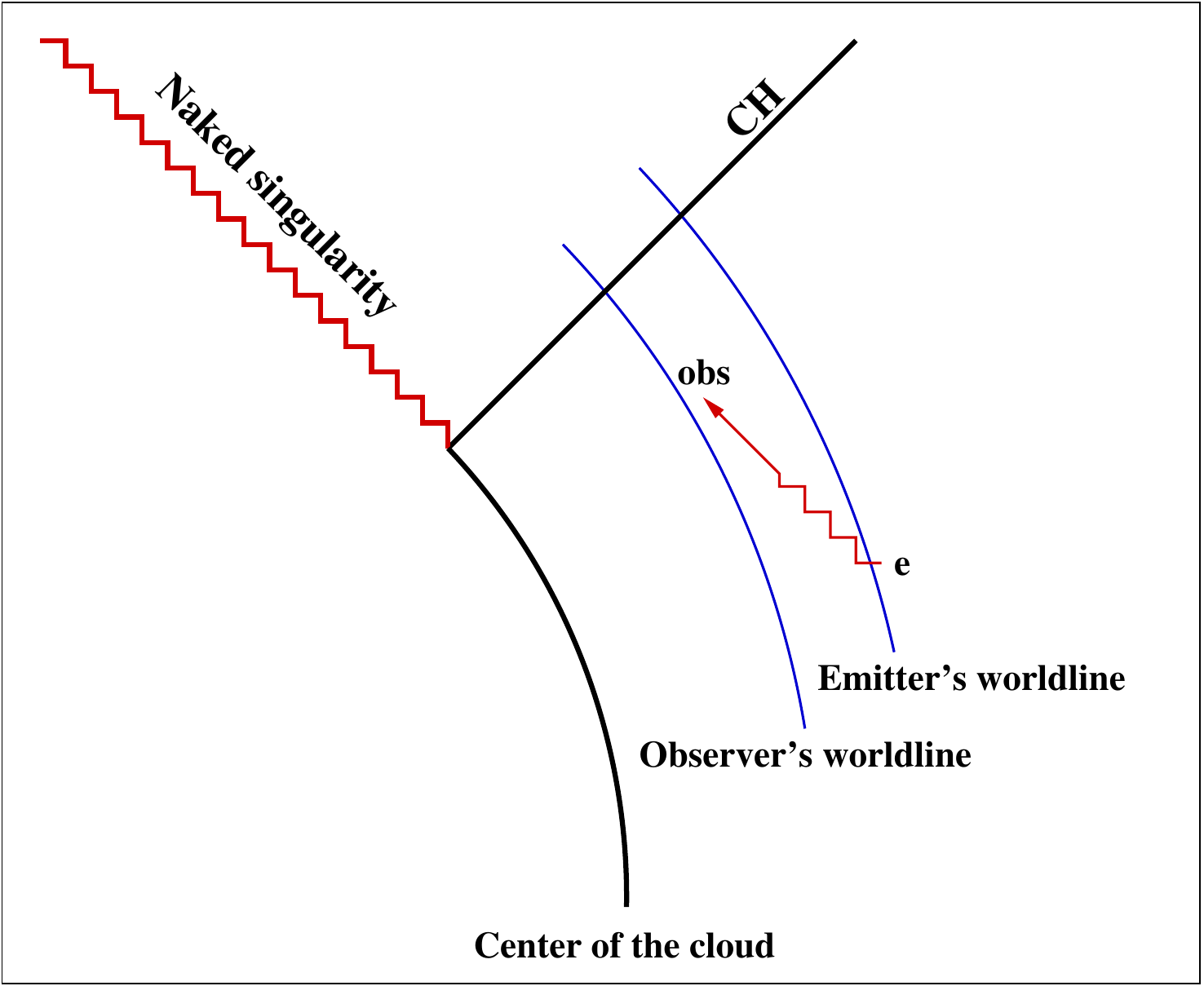}
\hspace{0.3cm}
\includegraphics[height=4.6cm]{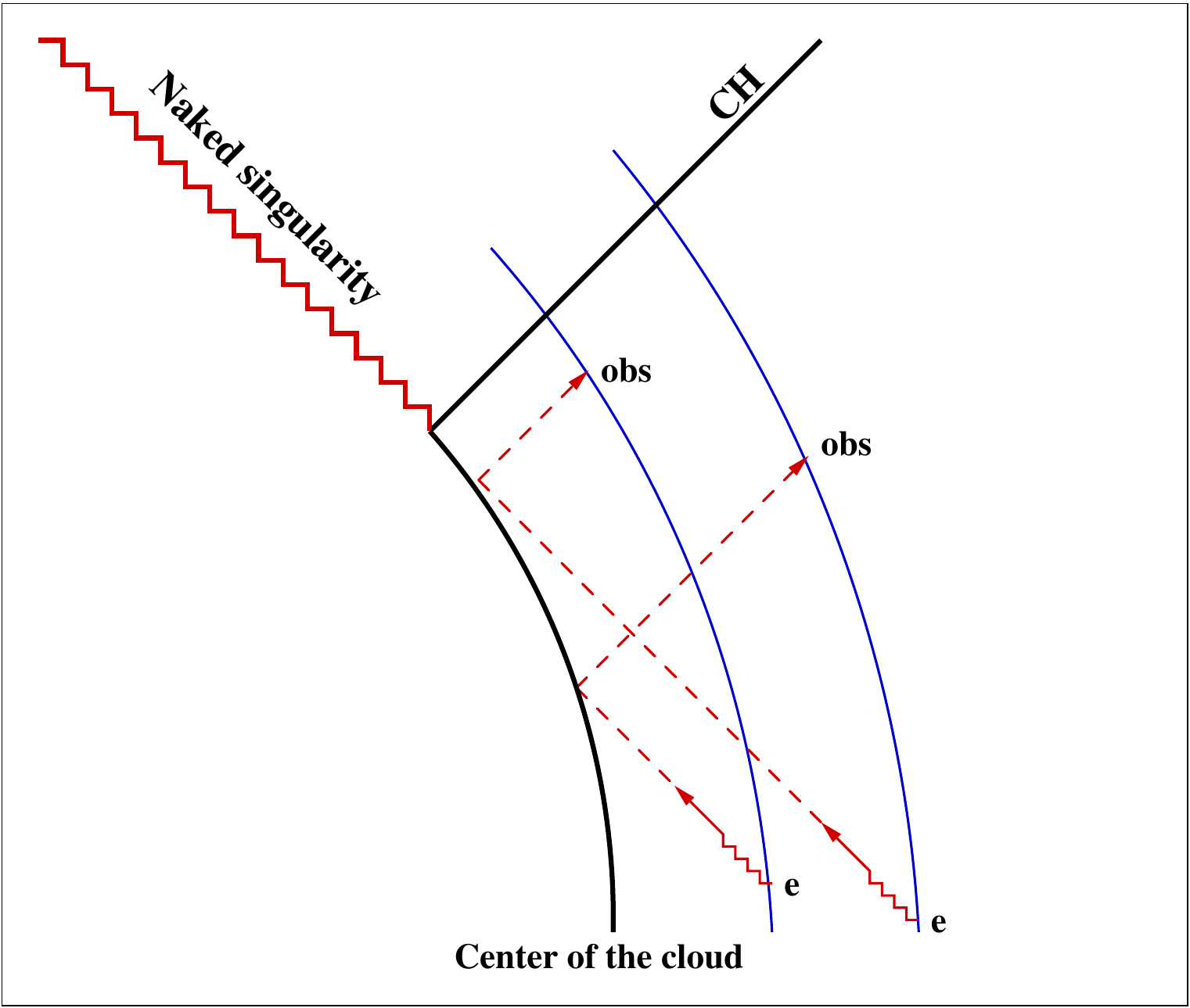}
\end{center}
\caption{\label{Fig:Blueshift} Left panel: A free-falling emitter ($e$) sends an outgoing radial null ray, which is received by a free-falling observer ($obs$). Middle panel: Same situation, but now ($e$) emits an ingoing radial null ray which is received by the observer ($obs$). Right panel: Similar situations, but now the observer is located at a point which is diametrically opposed to the emitter.}
\end{figure}

In this section we start by analyzing the behaviour of test fields on the background spacetime $(D,{\bf g})$ in the geometric optics approximation. In this limit, the propagation of test fields is described by null geodesics of $(D,{\bf g})$. This leads to the identification of the vector fields ${\bm \xi}\in {\cal X}(D)$ satisfying ${\bf g}({\bm \xi},{\bm \xi}) = 0$ and $\nabla_{\bm \xi} {\bm \xi} = 0$. Here, for simplicity, we restrict ourselves to radial null geodesics. Then, we consider an emitter $(e)$ which sends a high-frequency signal to an observer $(obs)$ and compute the frequency shift due to gravitational redshift effects. We assume that both the emitter and the observer are co-moving with the dust particles and thus have four-velocity ${\bf u}$, see figure~\ref{Fig:Blueshift} for possible scenarios. 

Under these assumptions the frequency shift is given by (see, for instance, Ref.~\cite{Straumann-Book})
\begin{equation}
\frac{\nu_{obs}}{\nu_e} = \frac{\left. {\bf g}({\bf u},{\bm \xi}) \right|_{obs}}
{\left. {\bf g}({\bf u},{\bm \xi})\right|_e}
\label{Eq:RedShift}
\end{equation}
with ${\bm \xi}$ the generator of the null rays. We are particularly interested in the case where the signal is emitted or received in a region that lies arbitrarily close to the central singularity, since an unboundedly large blueshift would indicate a large concentration of energy and thus an instability of the Cauchy horizon. In fact, there is also another interesting effect an unboundedly large blueshift could have.\footnote{We are thankful to M. Dafermos for explaining this point to us.} Namely, when the self-gravity of the test field is taken into account, a large blueshift along ingoing light rays could lead to the formation of a trapped surface in the self-consistent calculation. The presence of such a trapped surface would effectively censor the singularity~\cite{mD05b}, as occurs in the spherically symmetric collapse of a self-gravitating scalar field~\cite{dC87,dC94,dC99}.

In order to discuss the frequency shift, we first identify the radial vector fields ${\bm\xi}$ generating null geodesics. They must be parallel to either one of the two null vectors ${\bf k}$ and ${\bf l}$ defined at the end of the last section, so all that needs to be determined is the factor between ${\bm \xi}$ and ${\bf k}$ or ${\bm \xi}$ and ${\bf l}$.

\begin{lemma}
\label{Lem:Geodesics}
The affinely parametrized future-directed radial null geodesics of $(M,{\bf g})$ are generated by the vector fields ${\bm \xi}_+ = f_+ {\bf k}$ and ${\bm \xi}_- = f_- {\bf l}$, where the functions $f_\pm$ are determined by the advection equations
\begin{equation}
{\bf k}[f_+] = \frac{\dot{\gamma}}{\gamma} f_+,\qquad
{\bf l}[f_-] = \frac{\dot{\gamma}}{\gamma} f_-.
\end{equation}
\end{lemma}

\proof
With respect to the local coordinates $(\tau,R,\vartheta,\varphi)$ we have for ${\bm \xi} = {\bm \xi}_+$ or ${\bm \xi} = {\bm \xi}_-$:
\begin{equation}
0 = \nabla_{\bm\xi} \xi_b = \xi^a\nabla_a\xi_b = \xi^a(\nabla_a\xi_b - \nabla_b\xi_a)
 = \xi^a(\partial_a\xi_b - \partial_b\xi_a),
\label{Eq:xiGeod}
\end{equation}
where we have used the fact that ${\bm \xi}$ is null. Using the expressions
\begin{displaymath}
\xi_\pm^\tau = f_\pm,\quad \xi_\pm^R = \pm\gamma f_\pm,\quad 
\xi_\pm^\vartheta = \xi_\pm^\varphi = 0, 
\end{displaymath}
and
\begin{displaymath}
\xi_{\pm\tau} = -f_\pm,\quad \xi_{\pm R} = \pm\gamma^{-1} f_\pm,\quad
\xi_{\pm\vartheta} = \xi_{\pm\varphi} = 0,
\end{displaymath}
for the contravariant and covariant components of ${\bm \xi}_\pm$ we obtain form Eq.~(\ref{Eq:xiGeod}) the equation
\begin{equation}
\left( \frac{1}{\gamma}\frac{\partial}{\partial \tau} 
 \pm \frac{\partial}{\partial R} \right) f_\pm = \frac{\dot{\gamma}}{\gamma^2} f_\pm,
\label{Eq:fpm}
\end{equation}
which is equivalent to the claim.
\qed

An explicit solution representation for the functions $f_\pm$ can be obtained by the method of characteristics. For this, let $\tau_\pm(R)$ be a solution of
\begin{displaymath}
\frac{d}{dR}\tau_\pm(R) = \pm \frac{1}{\gamma(\tau_\pm(R),R)},
\end{displaymath}
describing an out- or ingoing radial null ray. Define $F_\pm(R) := f_\pm(\tau_\pm(R),R)$. Then, by virtue of Eq.~(\ref{Eq:fpm}), $F_\pm$ satisfy the ordinary differential equations
\begin{displaymath}
\frac{d}{dR} F_\pm(R) 
 = \pm \frac{\dot{\gamma}}{\gamma^2}(\tau_\pm(R),R) F_\pm(R).
\end{displaymath}
For the following we introduce the propagators
\begin{equation}
{\cal P}_\pm(R_2,R_1) := \exp\left[ \pm \int\limits_{R_1}^{R_2} \frac{\dot{\gamma}}{\gamma^2}(\tau_\pm(R),R) dR \right],
\label{Eq:Propagators}
\end{equation}
which propagate the values of $f_\pm$ along the radial light rays $(\tau_\pm(R),R)$ from $R = R_1$ to $R = R_2$. Using these propagators, Lemma~\ref{Lem:Geodesics} and the observation that ${\bf g}({\bf u},{\bf k}) = {\bf g}({\bf u},{\bf l}) = -1$, Eq.~(\ref{Eq:RedShift}) yields the following expression for the frequency shift due to gravitational redshift effects:
\begin{equation}
\frac{\nu_{obs}}{\nu_e} = {\cal P}_\pm(R_{obs},R_e),
\label{Eq:Blueshift}
\end{equation}
where the plus sign refers to the case where an outgoing signal is sent and the minus sign to the case of the ingoing signal.

Therefore, the question about the boundedness of the frequency shift is reduced to the analysis of the propagators~(\ref{Eq:Propagators}). We are particularly interested in their properties close to the central singularity. For this, we first note the following property of the integrand $\dot{\gamma}/\gamma^2$:

\begin{lemma}
\label{Lem:gamma}
For $\varepsilon > 0$ small enough there exists a constant $C_1 > 0$ such that on $D(\varepsilon)$,
\begin{equation}
0 < y\frac{\dot{\gamma}}{\gamma^2}  \leq C_1.
\end{equation}
\end{lemma}

\proof We start with the explicit computation of $\dot{\gamma}/\gamma^2$, based on Eq.~(\ref{Eq:gamma}) which can be rewritten as
\begin{displaymath}
\frac{1}{\gamma} = \frac{r'}{\sqrt{1 - R^2 q(R)^2 c(R)}}.
\end{displaymath}
Taking a partial derivative with respect $\tau$ on both sides of this equation gives
\begin{displaymath}
-\frac{\dot{\gamma}}{\gamma^2} = \frac{Q(R)}{\sqrt{c(R)}} \dot{r}',
\end{displaymath}
where the function $Q(R)$ was defined in the previous section. In order to evaluate $\dot{r}'$ we use the equation
\begin{displaymath}
\dot{r} = -\sqrt{2E(R) + \frac{2m(R)}{r}} = -\frac{R\sqrt{c(R)}}{y} 
\sqrt{1 - q(R)^2 y^2},
\end{displaymath}
which follows from the energy conservation law, see Eq.~(\ref{Eq:1DMechanical}). Together with
\begin{displaymath}
y' = \frac{R}{2y^2}\sqrt{1 - q(R)^2 y^2}\Lambda(y,R),
\end{displaymath}
we finally obtain
\begin{equation}
\frac{\dot{\gamma}}{\gamma^2} = \frac{Q(R)}{y} H(y,R),
\label{Eq:gamma_dot/gamma_sqrd}
\end{equation}
with the function
\begin{equation}
H(y,R) := \frac{1}{\sqrt{1-q(R)^2y^2}}
\left[ 1 + R\frac{c'(R)}{2c(R)} - q(R)^2 y^2\left( 1 + R\frac{c'(R)}{2c(R)} + R\frac{q'(R)}{q(R)} \right) \right]
 -\frac{1}{2}\frac{R^2}{y^3}\Lambda(y,R).
\end{equation}
The first term on the right-hand side is a continuous function of $(y,R)$ which converges to $1$ as $(y,R) \to (0,0)$. As for the second term, we use a similar estimate than Eq.~(\ref{Eq:R2y3Estimate}) which shows that inside region $D(\varepsilon)$,
\begin{displaymath}
0\leq \frac{R^2}{y^3}\Lambda(y,R) \leq \frac{4}{3}\frac{1+\varepsilon}{1-\varepsilon}.
\end{displaymath}
It follows that for small enough $\varepsilon > 0$, $0 < H(y,R) < 4/3$ for all $(y,R)\in D(\varepsilon)$. Since $Q$ is strictly positive and continuous the lemma follows.
\qed

A first important consequence of Lemma~\ref{Lem:gamma} is the positivity of ${\cal P}_+(R_{obs},R_e)$ for small enough $R_{obs} > R_e > 0$ and the positivity of ${\cal P}_-(R_{obs},R_e)$ for small enough $R_e > R_{obs} > 0$. This implies that the photons with zero angular momentum, as observed by co-moving observers close to the singularity, are \emph{blueshifted}. For the ingoing case, this might be expected since the photons move towards the central singularity. However, for the outgoing case the fact that the photons are blueshifted and not redshifted might come as a surprise since in this case the photons move away from the singularity. In fact, it was recently shown~\cite{lKdMcB13} for the marginally bound case that the blueshift of photons which are emitted by the dust particles in the interior of the cloud can be larger than their redshift in the exterior region so that in this case, a distant observer measures a blueshift.

In order to clarify why outgoing photons are blueshifted close to the singularity, we consider the compactness ratio
\begin{equation}
C(\tau,R) := \frac{2m(R)}{r(\tau,R)}
\end{equation}
as a measure for the strength of the gravitational field at $(\tau,R)$. Its variation along outgoing null geodesics is given by
\begin{displaymath}
k[C] = -\frac{2m}{r^2} k[r] + \frac{\gamma}{r} 2m'.
\end{displaymath}
As long as the null ray lies outside the apparent horizon, $k[r] > 0$, and the first term is negative. Since we assume positive mass density, the second term is positive inside the cloud. Outside the cloud, it vanishes and the photons are redshifted in the outside region, provided they arrive at the surface of the cloud before the horizon forms. In order to see which term dominates inside the cloud we use $2m = R^3 c$ and $r = Ry^2$ and rewrite
\begin{equation}
k[C] = \frac{2m}{r^2}\left[ -k[r] + \left( 3 + \frac{Rc'}{c} \right)\gamma y^2 \right].
\label{Eq:kC}
\end{equation}
Since
\begin{eqnarray*}
k[r] &=& \dot{r} + \gamma r' = -\frac{R\sqrt{c}}{y}\sqrt{1 - q^2 y^2} + \sqrt{1 - R^2 q^2 c},
\\
\gamma y^2 &=& \frac{\sqrt{1 - R^2 q^2 c}}{1 + \frac{R^2}{y^3}\sqrt{1 - q^2 y^2}\Lambda},
\end{eqnarray*}
we see that $k[r]$ and $\gamma y^2$ both converge to one when $R\to 0$ and $y$ is bounded away from zero. Consequently, the expression inside the square parenthesis on the right-hand side of Eq.~(\ref{Eq:kC}) converges to $2$ and since $2m/r^2 > 0$ it follows that $k[C] > 0$ is positive for outgoing radial light rays close to the center. Using the estimate~(\ref{Eq:R2y3Estimate}) it is not difficult either to show that $k[C]$ is positive in the interior of the region $D(\varepsilon)$ for sufficiently small $\varepsilon > 0$. In this sense the outgoing photons move towards a region of stronger gravity when they are close to the center of the cloud or close to the central singularity. This provides an explication for their blueshift in these regions.

A further important consequence of Lemma~\ref{Lem:gamma} concerns the \emph{boundedness} of the blueshift. We formulate this result in terms of the redshift factor
\begin{equation}
z := \frac{\nu_e}{\nu_{obs}} - 1 = {\cal P}_\pm(R_e,R_{obs}) - 1
 = \exp\left[ \mp \int\limits_{R_e}^{R_{obs}} \frac{\dot{\gamma}}{\gamma^2}(\tau_\pm(R),R) dR \right] - 1.
\label{Eq:zDef}
\end{equation}

\begin{proposition}
\label{Prop:Blueshift}
Consider the redshift factor $z$ along in- and outgoing radial geodesics in the region $D$ of spacetime. Then, $|z|$ is uniformly bounded inside the dust cloud, and $z$ is negative in a region of the form $D(\varepsilon)$ close to the central singularity.
\end{proposition}

\proof Let $D(\varepsilon)$ be such that the inequality in Lemma~\ref{Lem:gamma} holds. Since $\dot{\gamma}/\gamma^2$ is a smooth function on $D$ it is sufficient to prove that $|z|$ is uniformly bounded on $D(\varepsilon)$. For this, we use a similar estimate than Eq.~(\ref{Eq:R2y3Estimate}) in order to find
\begin{displaymath}
\frac{R^2}{y^3} \leq M^3 := \frac{4}{3\Lambda_0}\frac{1}{1-\varepsilon}
\end{displaymath}
on $D(\varepsilon)$. Therefore, it follows from Lemma~\ref{Lem:gamma} that inside $D(\varepsilon)$,
\begin{displaymath}
0 < \frac{\dot{\gamma}}{\gamma^2} \leq \frac{C_1}{y} \leq \frac{C_1 M}{R^{2/3}},
\end{displaymath}
implying that
\begin{displaymath}
0 < \int\limits_{R_e}^{R_{obs}} \frac{\dot{\gamma}}{\gamma^2} dR
Ê\leq C_1 M\int\limits_{R_e}^{R_{obs}} \frac{dR}{R^{2/3}} \leq 3C_1 M R_{obs}^{1/3} < \infty
\end{displaymath}
for $R_{obs} \geq R_e > 0$ sufficiently small. This proves that $z$ is uniformly bounded and negative on $D(\varepsilon)$. 
\qed

To summarize the results of this section, we have found that photons emitted and received by radial observers which are co-moving with the dust particles are blueshifted in a vicinity of the central singularity. The blueshift can be computed from the explicit expression in Eq.~(\ref{Eq:zDef}). Since $\dot{\gamma}/\gamma^2$ diverges as $1/y$ near the singularity, see Lemma~\ref{Lem:gamma}, the blueshift is more pronounced near the singularity, as expected. However, Proposition~\ref{Prop:Blueshift} shows that this blueshift is uniformly bounded and thus cannot be arbitrarily large even when the light ray passes arbitrarily close to the singularity. In particular, this implies that photons that are sent from a static observer outside the cloud, traverse the collapsing cloud and are received by another static observer outside the cloud cannot gain an arbitrarily large amount of energy, even if they pass close to the central singularity. In this respect, an observer outside the cloud will notice nothing particular as he reaches the Cauchy horizon.

\section{Boundedness of spherically symmetric effective test fields }
\label{Sec:SphSym}

In the last section we discussed the propagation of fields on the collapsing dust background in the geometric optics approximation. In this section we relax the high-frequency assumption leading to this approximation and consider instead the exact Cauchy evolution of massless and massive test scalar fields $\Phi$ from regular initial data on the Cauchy surface $\tau=0$. The dynamics of $\Phi$ is governed by an effective wave equation with potential $V$ for the rescaled field $\psi = r\Phi$ on the two-dimensional spacetime $(\tilde{M},\tilde{\bf g})$, where $\tilde{M} = \{ (\tau,R) : R \geq 0, 0\leq \tau < \tau_s(R) \}$, $\tilde{\bf g} = -d\tau^2 + dR^2/\gamma(\tau,R)^2$ and $r$ is the areal radius. Considering a solution $\psi$ on the domain of dependence $D$ of the initial surface $\tau=0$ satisfying suitable regularity conditions at $R=0$ we prove in Theorem~\ref{Thm:psiBoundedness} that $\psi$ can be continuously extended to the closure $\overline{D}$ of $D$, provided the potential $V$ is $L^1$-integrable on compact subsets of $(\tilde{M},\tilde{\bf g})$ and in a vicinity of the form $D(\varepsilon)$ of the naked singularity, see figure~\ref{Fig:D_epsilon}. This means that the field $\psi$ can be continuously extended to the Cauchy horizon, including the first singular point from which it emanates. In particular, this result implies that $\psi$ is uniformly bounded in the part $D_c$ of $D$ lying inside the cloud. For a spherically symmetric scalar field $\Phi$ of arbitrary mass we prove in Lemma~\ref{Lem:Int_V_bounded} that the integrability conditions on $V$ are satisfied, and thus we conclude that $r\Phi$ can be extended to the Cauchy horizon and is uniformly bounded on $D_c$.

When decomposed into spherically harmonics,
\begin{displaymath}
\Phi(\tau,R,\vartheta,\varphi) 
 = \frac{1}{r}\sum\limits_{\ell=0}^\infty\sum\limits_{m=-\ell}^\ell
\psi_{\ell m}(\tau,R) Y^{\ell m}(\vartheta,\varphi)
\end{displaymath}
the Klein-Gordon equation on an arbitrary spherically symmetric background spacetime reduces to a family of effective wave equations on $(\tilde{M},\tilde{\bf g})$ (see, for instance, Ref.~\cite{eCnOoS13})
\begin{equation}
\tilde{g}^{ab}\tilde{\nabla}_a\tilde{\nabla}_b\psi_{\ell m} = V_\ell\psi_{\ell m},
\end{equation}
with potential
\begin{equation}
V_\ell = \frac{\ell(\ell+1)}{r^2} + \frac{\tilde{g}^{ab}\tilde{\nabla}_a\tilde{\nabla}_b r}{r} 
 + \mu^2,
\label{Eq:EffectivePotential}
\end{equation}
where $\tilde{\nabla}$ denotes the Levi-Civita connection associated to $(\tilde{M},\tilde{\bf g})$ and $\mu$ is the inverse Compton length of the field $\Phi$. We stress that the effective potential $V_\ell$ diverges at the central singularity, even for $\ell=0$, see Eq.~(\ref{Eq:VEffective}) below. Therefore, one would naively expect that the scalar field diverges at the central singularity and that this divergence could propagate along the Cauchy horizon. Surprisingly, this is not the case as follows from the arguments below.

Our result can be stated as follows:

\begin{theorem}
\label{Thm:psiBoundedness}
Let $(\tilde{M},\tilde{\bf g})$ be the two-dimensional spacetime manifold $\tilde{M} = \{ (\tau,R) : R \geq 0, 0\leq \tau < \tau_s(R) \}$ with metric $\tilde{\bf g}$ obtained from the one in Eq.~(\ref{Eq:MetricSol}) by fixing the angular coordinates. Assume the conditions (i)--(viii) on the initial density and velocity profiles hold. Furthermore, suppose $V$ is a function on $(\tilde{M},\tilde{\bf g})$ which is locally integrable and such that for some $\varepsilon\in (0,1)$,
\begin{equation}
\int\limits_{D(\varepsilon)} |V(x)| \sqrt{|\det\tilde{g}(x)|} d^2 x < \infty.
\label{Eq:VIntegrability}
\end{equation}

Then, any solution $\psi$ of the effective wave equation
\begin{equation}
\tilde{g}^{ab}\tilde{\nabla}_a\tilde{\nabla}_b\psi = V\psi,
\label{Eq:EffectiveWave}
\end{equation}
on the domain of dependence $D$ belonging to smooth initial data on the initial surface $\tau=0$ and which satisfies the regularity condition $\psi(\tau,R=0) = 0$, $0\leq \tau < \tau_s(0)$, at the center can be continuously extended to the closure $\overline{D}$ of $D$. In particular, $\psi$ is uniformly bounded on the region $D_c$, that is, there exists a constant $C > 0$ such that for all $x\in D_c$
\begin{equation}
|\psi(x)| \leq C.
\end{equation}
\end{theorem}

{\bf Remark}: It is interesting to note that the extension principle of $\psi$ only requires the integrability condition~(\ref{Eq:VIntegrability}), and not any further conditions on the metric $\tilde{\bf g}$. The reason for this is that in terms of double null coordinates $(u,v)$ the two-metric has the form $\tilde{\bf g} = -\Omega^2 du dv$, where all singularities of $\tilde{\bf g}$ are contained in the positive conformal factor $\Omega$. In terms of these coordinates, Eq.~(\ref{Eq:EffectiveWave}) reads
\begin{equation}
-2\partial_u\partial_v\psi = \sqrt{|\det\tilde{g}(x)|} V(x)\psi,
\label{Eq:EffectiveWaveConfCoord}
\end{equation}
and we see that the singularity only enters the combination $\sqrt{|\det\tilde{g}(x)|} V(x)$.\\

\proof We first note that according to standard uniqueness and existence theorems, the solution $\psi$ must be at least continuous on $D$. The proof of Theorem~\ref{Thm:psiBoundedness} is based on the integration of Eq.~(\ref{Eq:EffectiveWaveConfCoord}) over a characteristic rectangle $E := [u_1,u_2]\times [v_1,v_2] \subset D$, which yields
\begin{equation}
-2 \left[ \psi(u_2,v_2) - \psi(u_1,v_2) - \psi(u_2,v_1) + \psi(u_1,v_1) \right] 
 = \int\limits_E V(x)\psi(x)\sqrt{|\det\tilde{g}(x)|} d^2x.
\label{Eq:ParallelIdentity}
\end{equation}

\begin{figure}[h!]
\begin{center}
\includegraphics[width=9cm]{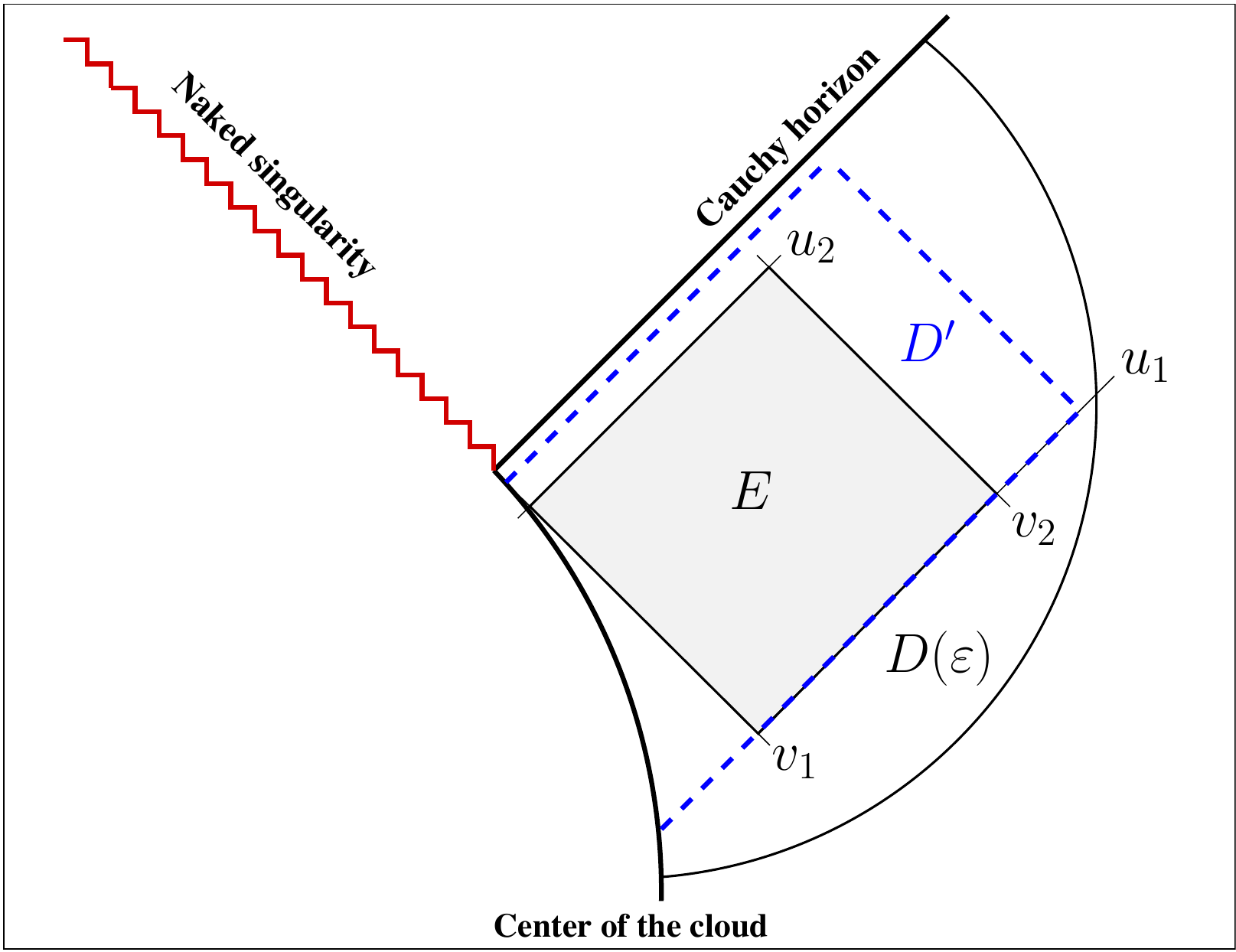}
\end{center}
\caption{\label{Fig:Domain_D} Illustration of the regions used in the proof of the extension principle.}
\end{figure}

We first show that $\psi$ is uniformly bounded inside the region $D(\varepsilon)$. For this, we choose $E$ such that the point $(u_2,v_1)$ lies on the center of the cloud (see figure~\ref{Fig:Domain_D}). Then, since $\psi = 0$ at the center, we obtain the implicit relation
\begin{equation}
\psi(u_2,v_2) = \psi(u_1,v_2) - \psi(u_1,v_1(u_2)) 
 - \frac{1}{2} \int\limits_E V(x)\psi(x)\sqrt{|\det\tilde{g}(x)|} d^2 x
\label{Eq:ParallelIdentityBis}
\end{equation}
for $\psi$, where $v_1(u_2)$ indicates that $v_1$ depends on $u_2$. This yields the estimate
\begin{equation}
|\psi(u_2,v_2)| \leq |\psi(u_1,v_2)| + |\psi(u_1,v_1(u_2))|
 + \frac{1}{2}\int\limits_E |V(x)| |\psi(x)| \sqrt{|\det\tilde{g}(x)|} d^2x.
\label{Eq:EstimateScal1}
\end{equation}

Next, we fix $\varepsilon > 0$ and $u_1$, and choose a closed subset $D'\subset D(\varepsilon)$ of the form depicted in figure~\ref{Fig:Domain_D}. Moreover, we introduce the quantities
\begin{eqnarray*}
A &:=& 2\sup\{  |\psi(u_1,v) | : \hbox{$v$ such that $(u_1,v)\in D'$} \},\\
B &:=& \frac{1}{2} \int\limits_{D(\varepsilon)} |V(x)| \sqrt{|\det\tilde{g}(x)|}  d^2x,
\end{eqnarray*}
which are finite according to the hypothesis of the theorem. With this notation the estimate~(\ref{Eq:EstimateScal1}) implies that
\begin{equation}
|\psi(u_2,v_2)| \leq A + B\sup\limits_{(u,v)\in D'} |\psi(u,v)|
\label{Eq:EstimateScal2}
\end{equation}
for all $(u_2,v_2)\in D'$. Taking the supremum over $(u_2,v_2)\in D'$ on both sides yields the simple inequality
\begin{displaymath}
x \leq A + Bx,\qquad x := \sup\limits_{(u,v)\in D'} |\psi(u,v)|.
\end{displaymath}
Finally, since $B$ is finite, we can make it arbitrarily small by choosing $\varepsilon > 0$ sufficiently small. (This follows from the dominated convergence theorem, see for instance~\cite{Royden-Book}.) In particular, if $B < 1$ we conclude that
\begin{displaymath}
x\leq \frac{A}{1 - B} < \infty,
\end{displaymath}
which proves that $\psi$ is uniformly bounded on the union of all closed subsets $D'\subset D(\varepsilon)$ of the form depicted in figure~\ref{Fig:Domain_D} with fixed $u_1$. Since according to the assumptions of the theorem $|V|$ is integrable on any compact subset $K\subset \tilde{M}$  we can use similar arguments and conclude that $\psi$ is bounded on any subset of the form $K\cap D$.

Next, we prove the extension result. For this, let $(u_c,v_c)$ be a point on the Cauchy horizon, and let $(u^{(n)},v^{(n)})$ be a sequence in $D$ which converges to $(u_c,v_c)$. Evaluating the implicit relation~(\ref{Eq:ParallelIdentityBis}) at $(u_2,v_2) = (u^{(k)},v^{(k)})$ and $(u_2,v_2) = (u^{(n)},v^{(n)})$ and taking the difference we obtain the estimate
\begin{eqnarray}
| \psi(u^{(k)},v^{(k)}) - \psi(u^{(n)},v^{(n)}) | &\leq&
 | \psi(u_1,v^{(k)}) - \psi(u_1,v^{(n)}) | 
 + | \psi(u_1,v_1(u^{(k)})) - \psi(u_1,v_1(u^{(n)})) |
\nonumber\\
 &+& M\int\limits_{E^{(k)}\bigtriangleup E^{(n)}} |V(x)| \sqrt{|\det\tilde{g}(x)|} d^2x,
\label{Eq:EstimateScal3}
\end{eqnarray}
where $E^{(k)}\bigtriangleup E^{(n)}$ denotes the symmetric difference of $E^{(k)}$ and $E^{(n)}$, that is, the set of points which are in either of the two sets but not in their intersection. In deriving this estimate we have used the previous boundedness result on $\psi$. Since $\psi$ is continuous on $D$ and because of the integrability condition on $V$, it follows that the right-hand side of Eq.~(\ref{Eq:EstimateScal3}) vanishes in the limit $k,n\to\infty$. Consequently, it follows from Eq.~(\ref{Eq:EstimateScal3}) that $\psi(u^{(n)},v^{(n)})$ is a Cauchy sequence in $\Real$, and the limit
\begin{displaymath}
\lim\limits_{n\to\infty} \psi(u^{(n)},v^{(n)})
\end{displaymath}
exists in $\Real$. The limit is independent of the sequence that converges to $(u_c,v_c)$, because if $(\tilde{u}^{(n)},\tilde{v}^{(n)})$ is another sequence that converges to $(u_c,v_c)$, it follows by replacing $(u^{(k)},v^{(k)})$ with $(\tilde{u}^{(n)},\tilde{v}^{(n)})$ in Eq.~(\ref{Eq:EstimateScal3}) and by taking the limit $n\to\infty$ that
\begin{displaymath}
\lim\limits_{n\to\infty} \psi(\tilde{u}^{(n)},\tilde{v}^{(n)}) 
 = \lim\limits_{n\to\infty} \psi(u^{(n)},v^{(n)}) =: \psi(u_c,v_c).
\end{displaymath}
This defines a continuous extension of $\psi$ on $\overline{D}$. In particular, $\psi$ vanishes at the central naked singularity since $\psi$ is zero at the center.
\qed

The next lemma shows that the integrability conditions on $V_\ell$ are satisfied in the spherically symmetric case $\ell=0$.

\begin{lemma}
\label{Lem:Int_V_bounded}
For $\ell = 0$ the effective potential $V_0$ defined in Eq.~(\ref{Eq:EffectivePotential}) is locally integrable, and for all $\varepsilon\in (0,1)$ it satisfies
\begin{equation}
\int\limits_{D(\varepsilon)} |V_0(x) | \sqrt{|\det\tilde{g}(x)|}  d^2x < \infty.
\label{Eq:B_lt_infty}
\end{equation}
\end{lemma}

\proof In terms of the coordinates $(y,R)$ the effective potential~(\ref{Eq:EffectivePotential}) is given by
\begin{equation}
V_\ell(y,R) = \frac{\ell(\ell+1)}{R^2 y^4} + \frac{c(R)}{y^6} 
 - \frac{Rc'(R) + 3c(R)}{2y^4 r'} + \mu^2,
\label{Eq:VEffective}
\end{equation}
and the weight in the volume element belonging to the two-metric is
\begin{displaymath}
\sqrt{|\det\tilde{g}(y,R)|} = \frac{1}{\sqrt{c(R)}\sqrt{1 - q(R)^2 y^2}}
\frac{2y^2 r'}{\sqrt{1-R^2 q(R)^2 c(R)}}.
\end{displaymath}
Therefore, the lemma reduces to the verification of the integrability of the function
\begin{equation}
V_\ell\sqrt{|\det\tilde{g}|} = \frac{1}{\sqrt{c}\sqrt{1 - q^2 y^2} \sqrt{1-R^2 q^2 c}}\left[
\frac{2\ell(\ell+1)A}{R^2} + \frac{2c A}{y^2} - \frac{Rc' + 3c}{y^2} + \mu^2 y^4 A \right],
\label{Eq:Vldet}
\end{equation}
where we have introduced the function
\begin{displaymath}
A(y,R) := \frac{r'}{y^2} = 1 + \frac{R^2}{y^3}\sqrt{1 - q(R)^2 y^2}\Lambda(y,R).
\end{displaymath}

First, we note that this function is continuous on $D$, implying the local integrability of $V_0$. Next, we note that the function $A(y,R)$ is bounded in $D(\varepsilon)$ according to the estimate~(\ref{Eq:y2Estimat}). Since the factor in front of the square parenthesis in Eq.~(\ref{Eq:Vldet}) is bounded, it follows from this that exists a constant $k > 0$ such that
\begin{displaymath}
| V_0 |\sqrt{|\det\tilde{g}|} \leq \frac{k}{y^2}
\end{displaymath}
on $D(\varepsilon)$. Now the lemma is a consequence of the integrability of the function $1/y^2$ over the region $D(\varepsilon)$. In order to show this, recall the definition~(\ref{Eq:DepsDef}) of $D(\varepsilon)$, which implies in particular that
$y_{CH}(R)^3 \geq \frac{3\Lambda_0}{4}(1-\varepsilon) R^2$ for all $(y,R)\in D(\varepsilon)$. Therefore,
\begin{eqnarray*}
\int\limits_{D(\varepsilon)} \frac{1}{y^2} dy dR
&=& \int\limits_0^{R(\varepsilon)} \left( 
 \int\limits_{y_{CH}(R)}^{y(\varepsilon)} \frac{dy}{y^2} \right) dR\\
&=& \int\limits_0^{R(\varepsilon)}
 \left( \frac{1}{y_{CH}(R)} - \frac{1}{y(\varepsilon)} \right) dR\\
&\leq& \left( \frac{4}{3\Lambda_0(1-\varepsilon)} \right)^{1/3}
  \int\limits_0^{R(\varepsilon)}\frac{dR}{R^{2/3}} - \frac{R(\varepsilon)}{y(\varepsilon)}\\
&=& 3\left( \frac{4}{3\Lambda_0(1-\varepsilon)} \right)^{1/3} R(\varepsilon)^{1/3}
 - \frac{R(\varepsilon)}{y(\varepsilon)} < \infty,
\end{eqnarray*}
which is finite.

In contrast, the function $1/R^2$ is not integrable over $D(\varepsilon)$ and consequently, the statement of the lemma cannot be generalized to the case $\ell > 0$. In this case, bounds have to be obtained via energy estimates.
\qed

As a consequence of Theorem~\ref{Thm:psiBoundedness} and Lemma~\ref{Lem:Int_V_bounded} we have:

\begin{theorem}
Consider a smooth, spherically symmetric solution $\Phi$ of the Klein-Gordon equation on the domain $D$ with smooth and bounded initial data for $\Phi$ and $\dot{\Phi}$ on the initial surface $\tau = 0$. Then, there exists a continuous extension of $\Phi$ on $\overline{D}\setminus \{(0,0)\}$ and there is a constant $C$ such that 
\begin{equation}
|\Phi(x)| \leq \frac{C}{r}
\label{Eq:PhiBound1}
\end{equation}
for all $x\in D_c$ in the interior of the cloud.
\end{theorem}

It should be stressed that this bound does not guarantee the finiteness of $\Phi$ at the naked central singularity, since it allows $\Phi$ to diverge as $r\to 0$. However, the importance of this result is that it rules out an infinite field propagating along the Cauchy horizon. In other words, even if $\Phi$ were to diverge at the central singularity, this divergence could not propagate along the Cauchy horizon.

\section{Discussion}
\label{Sec:Conclusions}

In this work we have analyzed physical effects that occur in the vicinity of a central singularity arising in the TB dust collapse. The initial data for the density and velocity profiles satisfy a set of reasonable assumptions, but are generic otherwise. In particular, we do not restrict ourselves to the marginally bound case nor to the self-similar collapse. The first effect we analyzed is the gravitational redshift along radial light rays, as measured by two observers which are co-moving with the dust particles. We showed that in the vicinity of the singularity there is a blueshift along in- and outgoing rays. As we explained, this effect is due to the collapse of the cloud, which implies that outgoing photons move towards a region of stronger gravity, and thus they gain energy. However, we also showed that the blueshift is uniformly bounded close to the singularity and the Cauchy horizon. An infinite blueshift would have indicated an instability of the Cauchy horizon, as is the case at inner horizons of black holes~\cite{mSrP73,ePwI90,mD05}. The fact that we could exclude an infinite blueshift does not, of course, prove that the Cauchy horizon is stable, but it indicates that its stability properties might be much more subtle than in the case of inner horizons.

Motivated by the results in the geometric optic approximation, we considered a spherically symmetric test scalar field $\Phi$ propagating on the background geometry of the TB dust collapse. Our main result is the extension principle proven in Theorem~\ref{Thm:psiBoundedness} which states that the rescaled field $r\Phi$ can be continuously extended to the Cauchy horizon. This result has several interesting consequences. First, it implies that it is possible to extend the scalar field continuously beyond the Cauchy horizon, provided appropriate data is given on an ingoing characteristic surface excising the null portion of the singularity. Second, the extension principle immediately implies that $\Phi$ is bounded by a constant divided by the areal radius $r$, close to the singularity. An important consequence of this bound is that it shows that even if $\Phi$ diverged at the central singularity, this divergence could not propagate along the Cauchy horizon. Although the techniques used to prove this bound cannot be extended to non-spherical fields, such as electromagnetic or linearized gravitational waves, this result is interesting in view of the results reported by numerical studies for gravitational perturbations, where it was found that a naked singularity could not be an important source of radiation~\cite{hItHkN98,hItHkN99,hItHkN00}.

In a follow up work we derive uniform energy bounds for more general fields propagating on the dust collapse background. In particular, we analyze the stability of linearized gravitational and fluid perturbations of the TB model.


\acknowledgments
We thank Mihalis Dafermos and Thomas Zannias for fruitful and stimulating discussions. This work was supported in part by CONACyT Grants No. 46521 and 101353 and by a CIC Grant to Universidad Michoacana.

\bibliographystyle{unsrt}
\bibliography{refs_collapse}

\end{document}